\documentclass[journal]{vgtc}                     

\onlineid{0}

\vgtccategory{Research}

\vgtcpapertype{please specify}

\title{Charting the Moral Universe: \\Capturing Virtues and Values of Data Visualization Practice}

\author{%
  \authororcid{Chloe Hudson Prock}{0009-0003-6445-5864},
  \authororcid{Enrico Bertini}{0000-0002-9932-0551},
  \authororcid{Michael Correll}{0000-0001-7902-3907}
}

\authorfooter{
  \item
  	Northeastern University, \{hudsonprock.c,e.bertini,m.correll\}@northeastern.edu
}

\abstract{%
What do we value in our visualizations, and in the people who design them?
Despite a growing body of work on critical data visualization, the conception of what it is to do ethical data visualization work can often be narrow (for instance, holding that our ethical duties are discharged merely by avoiding overtly lying or manipulating data), or entangled with potentially problematic implicit value structures (such as the assumption of the objectivity and neutrality of data, and so the designer's role being merely the passive conveying of numbers as efficiently as possible).
Yet, what it means to act ethically in data visualization is broad and multifaceted, and the virtues to which we should aspire as data visualization researchers and designers are worth explicating.
We conducted an interview study with a broad spectrum of 20 experienced data visualization researchers, practitioners, and data artists to solicit their values and ethical considerations around doing visualization work.
We report on a list of 68 values, organized into nine virtue clusters, that we encountered in our interviews. These virtues and values together describe a diverse space of matters of care and concern in data visualization: from the unease around the best use of visualization as a tool for persuasion, to the tightrope that visualization practitioners often walk between their professional responsibilities and their personal moral commitments.
The virtues themselves, as well as our interviewees' reflections on ethical practice, offer practitioners, researchers, and educators in data visualization a richer vocabulary for ethical reflection and provide a broader foundation for considering and applying visualization ethics.

}


\teaser{
\centering
 \includegraphics[width=\textwidth]{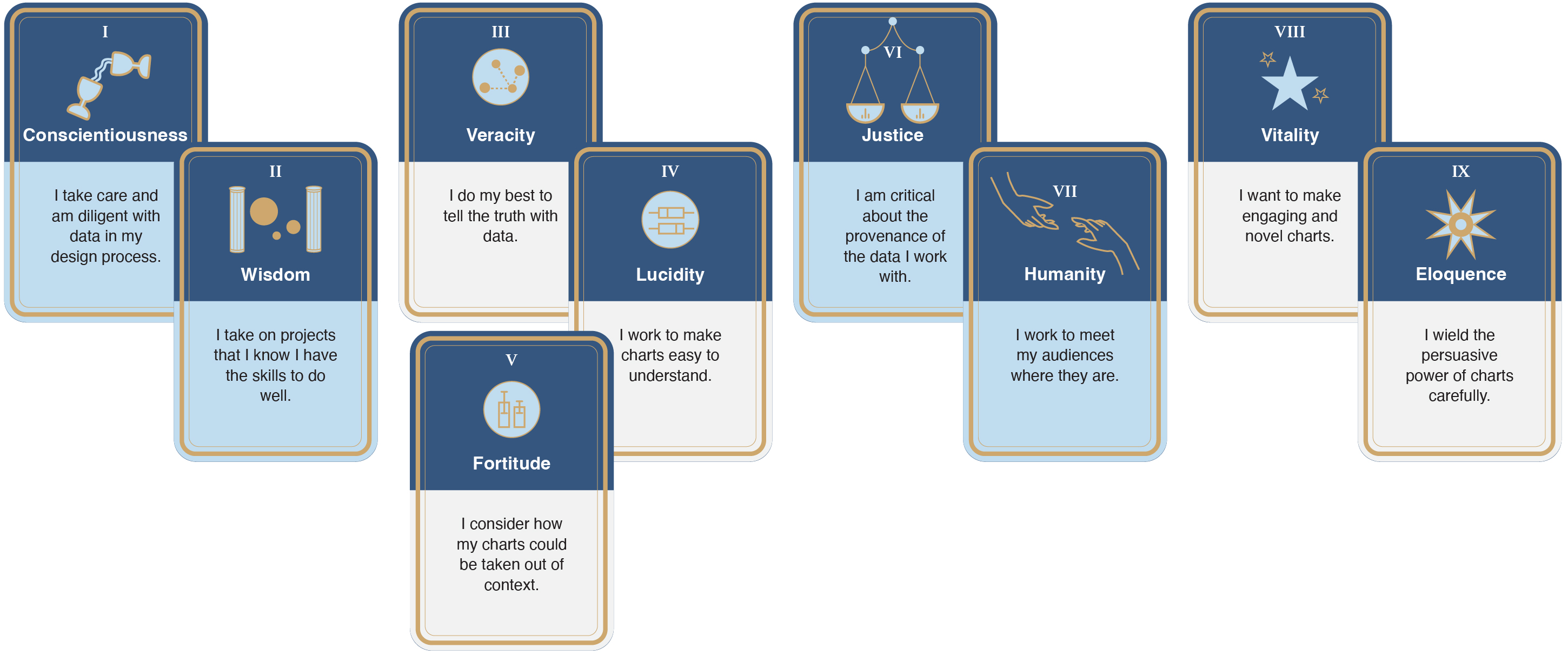}
 \caption{Symbolic representations of our nine \textit{virtue clusters}: semantically related groupings of 68 virtues and values, which are aspects of a what make a ``good'' visualization, or a good visualization designer.}
  \label{fig:teaser}
}

\graphicspath{{figs/}{figures/}{pictures/}{images/}{./}} 

\usepackage{tabu}                      
\usepackage{booktabs}                  
\usepackage{lipsum}                    
\usepackage{mwe}                       

\usepackage{graphicx}

\usepackage{subcaption}

\usepackage{soul} 

\usepackage{mathptmx}                  

\begin{document}



\firstsection{Introduction}

\maketitle
Journalists, researchers, teachers, citizen-scientists, data artists, and designers all have to navigate a complex ethical landscape when they perform their visualization work. Yet, past work on visualization ethics can focus heavily on negative examples: avoiding ``lies''~\cite{cairo2014graphics}, being ``misleading''~\cite{lisnic_misleading_2023}, ``misinforming''~\cite{lo2022misinformed}, or engaging in ``black hat''~\cite{correll2017black} practices. While the space of ``visualization for villainy''~\cite{mcnutt_visualization_2021} is broad and worrying, this focus on intentional bad behavior has several negative repercussions for cultivating ethical practices.

Assuming that the only harms in visualization arise from intentionally malicious actors encourages designers---unlikely to self-identify as malicious---to believe that they have satisfied all of their ethical obligations, limiting curiosity \textit{about} or moral ownership \textit{of} what can be written off as ``unintended consequences''~\cite{parvin2020unintended} of their work. One form of this neglect is the ``fatal premise'' in tech ethics~\cite{myers2012responsible}: ``Evil is done by evil people; I am not an evil person and therefore I cannot do evil.''  In visualization, the fatal premise can lead to the (often problematic, or at least ambiguous~\cite{kostelnick_visual_2008}) assumption that the ethical goals of the designer have been discharged once data have been ``clearly'' presented, and ignore the ``higher order of care''~\cite{gotterbarn2021being} required of those working with impactful and socially relevant data~\cite{d20206,schwabish_no_nodate}. There is more to the ethical design of visualizations than simply avoiding lying outright, in the same way that there is more to being a good person than avoiding the commission of capital crimes. 

Rather, ethical data visualization involves matters of judgment, taste, and care (that, admittedly, can be supplemented by best practices, recommendations, and design defaults). In short, \textit{phronesis}~\cite{noel1999varieties}: moral skill and practical wisdom brought about by personal development and experience. Building this expertise is connected with the cultivation and embodiment of \textit{virtues and values}. While we acknowledge that different ethical theories can have different accounts of what is meant by a ``virtue''~\cite{sep-ethics-virtue}, for this paper we use the terminology of \textit{virtues} to refer to ethical ``matters of concern''~\cite{ozkaramanli2017me} that can range from  \textit{object virtues} of what we find makes a good visualization (say, a chart that is efficient, effective, and aesthetically appealing), or the \textit{personal virtues} of what makes a good visualization designer (say, a designer who is honest, conscientious, and forthright). A ``value'', by comparison, refers to a concern that may or may not have an ethical component. This definition follows Friedman et al., who define a value as something that ``a person or group of people consider important in life''~\cite{friedman_value_2013}. As an example, a visualization designer may \textit{value} creating a visualization using their favorite color, but would be no more or less \textit{virtuous} for choosing to do so.

Virtues and values in visualization work are often tacit, and can rest on problematic assumptions~\cite{saharan2026critical} (for instance, that a good visualization is inherently universally understandable, or that data are inherently neutral). We therefore build on prior work in critical data visualization~\cite{correll_ethical_2019,dork_critical_2013}) to surface virtues and values with greater intentionality. We do so with the intention of both broadening our horizons on what it means to do ethical work, and building up a richer space of positive examples that is reflective of our community practices.

In this paper, we use a set of 20 semi-structured interviews with a variety of experienced data visualization researchers, journalists, designers, and educators to explore the virtues and values in circulation within visualization research, pedagogy, and practice. We present a resulting list of 68 values, grouped into nine virtue clusters, characterizing a broad set of ethical considerations. Our list of virtues and values is aimed at designers, design students, and evaluators of statistical graphics and data visualizations to expand ideas of ``good'' or ``bad'' visualization beyond merely avoiding deception. We do \textit{not} present an exhaustive list of ethical concerns or definitive guides to right action or an unimpeachable guide to ethical practice (if such a document were even possible). We argue against inflexible rules of ethical visualization design, or ethics as a box-ticking exercise of binary compliance with a list of rules, and instead advocate for designers to focus on personal growth and cultivation of multiple virtues, and the articulation of their personal values in their own work. As a result of our analyses, we find that ethical matters of concern are often in \textit{tension}, with even experienced visualization designers forced into ethical areas of discomfort; torn, for instance, between wanting to persuade their audience versus avoiding the appearance of bias or partiality, or between satisfying the client or remaining true to their personal beliefs.

\section{Background}
There are many ethical theories, and diversity within ethical perspectives that have surfaced in academic HCI and visualization venues~\cite{nunes_vilaza_scoping_2022}. A full review of normative ethical theories, or even just virtue ethics specifically, is outside the scope of this work, but we touch on a few factors relevant to our specific goals. While virtue ethics, which ``emphasizes the virtues, or moral character, in contrast to approaches emphasizing duties or rules (deontology) or the consequences of actions (consequentialism)''~\cite{sep-ethics-virtue} is of particular relevance to our framing, many normative ethical theories can and do make use of the language of virtues and values~\cite{sep-ethics-virtue}. We focus on virtues rather than other potential ethical features (for instance, Van Wijk's~\cite{van2005value} calculation of the ``value of visualization'' in terms of the utility of its insights versus the cost of its construction, which is a utilitarian formulation of good visualization design) for a variety of reasons. First, we do not wish for designers to feel that the process of ethically creating data visualizations is a rote matter of learning a set of rules (if such a set could even be formed; we note how even recommendations grounded in prior empirical work in visualization and graphical perception often overlap and contradict~\cite{zeng2023too}), but instead to think of their work in terms of career-long or life-long habits of critical thinking and applied moral wisdom (phronesis). Second, we find the language of virtues and values generative, allowing designers to consider a full spectrum of concerns, the balancing and consideration of which involve cultivating phronesis at every stage of design: Conwill et al.~\cite{conwill2025design}, for instance, find virtues to be productive for ideating about specific ``design patterns'' that can apply across technical domains.

The concerns motivating our work are not unique to visualization: scholars across data science and AI ethics have similarly worked to surface and systematize the values and principles at stake in technical practice ~\cite{floridi_what_2016, khan_ethics_2022}. Value Sensitive Design ~\cite{friedman_value-sensitive_1996, friedman_value_2013} offers an influential theory and method of accounting for human values in a comprehensive way throughout the design process. This work complements Value Sensitive Design's focus on how values are embedded in artifacts and design processes by descriptively surfacing the values that practitioners already hold and navigate in practice. HCI researchers have also grappled with the situational and context-dependent nature of ethical practice ~\cite{munteanu_situational_2015}, a challenge equally present in visualization design work. 

Our work draws on the burgeoning area of critical visualization---``the practice of examining and representing data with an awareness of the cultural, social, and ethical implications.''~\cite{panagiotidou2025critical} Foregrounding ethical considerations often involves surfacing or making explicit one's values, as in Do\"rk et al.'s~\cite{dork_critical_2013} call ``to develop a critical approach that examines the intentions behind visualizations and possible implications of their use,'' assessing strategies for designing visualizations that can support principles like ``disclosure, plurality, contingency, and empowerment.'' Correll~\cite{correll_ethical_2019}, using a similar virtue-based lens as this work, likewise charges visualization researchers and designers to ``make the invisible visible,'' ``collect data with empathy,'' and ``challenge structures of power.'' Critical visualization has already proven to be a transformative tool for enriching our conception of the field, provoking deeper reflection, and reducing harm: works like Lupi's~\cite{lupi2017data} ``data humanism'' manifesto and Klein \& D'Ignazio's ``data feminism''~\cite{d20206} project have wide currency in the community, and ``critical data visualization'' courses appear on university syllabi~\cite{mcnutt2025teaching}. While we build on the \textit{prescriptive} aspects of these critical projects, we are also interested in the \textit{descriptive} aspects of the ethical space: what values and virtues are widespread in the field as it is currently constituted?

In this descriptive sense, we connect with recent work to make explicit and prominent aspects of data visualization practice that are inescapable in visualization work, but often excluded from description in visualization research. For instance, Akbaba et al.~\cite{akbaba_entanglements_2024} point to how visualizations as research artifacts are epistemically entangled with interconnected structures of power and social relations. Likewise, Berret \& Munzner~\cite{berret2024iceberg} point to how the visible portion of the ``iceberg'' that is the concept of schemas belies implicit value and power structures that can shape the outcomes of visualization design work just as strongly as the more overtly considered elements of visual design and task analysis. We hold that ethical values are another such example of an under-articulated but omnipresent concept in data visualization work: we bring to our work many diverse value systems which are rarely, if ever, made explicit when conceptualizing or describing our work.

A concern when values are \textit{not} appropriately considered or surfaced is that these tacit values nonetheless exert undue and unconsidered force on visualization pedagogy, research, and practice. For instance, Drucker~\cite{drucker2011humanities} calls data visualizations an ``intellectual Trojan horse'' that bring along unearned assumptions of objectivity. Likewise, Houser~\cite{houser2014aesthetics} suggests that conceptualizations and rhetorics of information visualization are subject to ``contamination by transparency''~\cite{houser2014aesthetics}  that ``cause the obscurity and even trickery of data to recede from view.'' Kostelnick~\cite{kostelnick_visual_2008} likewise mentions the ``conundrum of clarity'' where great stock is set by the extent to which data visualizations are ``clear,'' despite the problematic and often contradictory meanings of the term. Recent work that most directly surfaces the danger of having the value structure underlying visualization work be tacit rather than explicit is Saharan et al.~\cite{saharan2026critical}: through a textual and historical analysis of prominent works in visualization pedagogy, they find the unquestioned assumptions that visualizations should embody ``universality, objectivity, and efficiency'' limit the diversity, reach, and nuance of visualization work. Our work shares the goal of making values and assumptions explicit (and so subject to potential exploration or critique), although we rely on interviews and seek to capture a wider collection of virtues and values. We do so in order to better understand not just the ethical \textit{aspirations} of the field, but a more \textit{descriptive} notion of ethical practices and matters of concern that are on the minds of thought leaders and high-impact members of the field. This descriptive aspect is important: in their meta-ethical critique of cartographic codes of ethics, Field \& Muehlenhaus~\cite{field2024ethics} find that value-based ethics codes can fail by being mere empty statements of general goodwill, or by not accounting for the realities of everyday work in the data profession.

\section{Methods}

In order to generate meaningful guidance for acting virtuously when designing statistical graphics and data visualizations, we require a set of matters of concern: the virtues and values associated with a ``good'' data visualization, or being a ``good'' data visualization designer. We do not suppose any list of virtues is likely to be exhaustive or objective. We follow Chappell~\cite{chappell2015lists}, analyzing Aristotle's methods in \textit{Nicomachean Ethics}, by claiming that a productive way of generating lists of virtues is to examine virtues in common circulation, and then refine and interrogate those virtues with respect to their mutual coherence and intelligibility. We further base our methods on projects like ``The Virtues Project''~\cite{newstead2020virtues}, a collation of ``virtue words'' found in common cross-cultural circulation. Even the brief lists and descriptions of virtues in The Virtues Project have shown both empirical~\cite{newstead2020evaluating} and theoretical~\cite{annas2004being} promise as tools for developing practical ethics. Ultimately, the goal of such lists is not purely descriptive, but often has a prescriptive aspect: ``not so much attempts to describe a reality that is already there, as to summon a reality into being by sketching an ideal and exhorting one's hearers to live up to it''~\cite{chappell2015lists}.

We generate our list of virtues through a series of semi-structured interviews with 20 experts across visualization research and practice. We performed a thematic analysis on the resulting interview transcripts to identify both explicit value words as well as statements that point to values or virtues. This analysis resulted in a list of 68 values, which we semantically sort into nine  ``virtue clusters''~\cite{olszewska2021ieee} of related matters of concern and discuss in four groups. While not exhaustive, we intend for this list to be expansive and generative--- reflective both of real matters of concern in data visualization, but also as ways to uncover un- or under-explored ethical territory in critical data visualization.

\subsection{Participants}

\begin{figure}[h]
    \centering
    \includegraphics[width=0.5\textwidth]{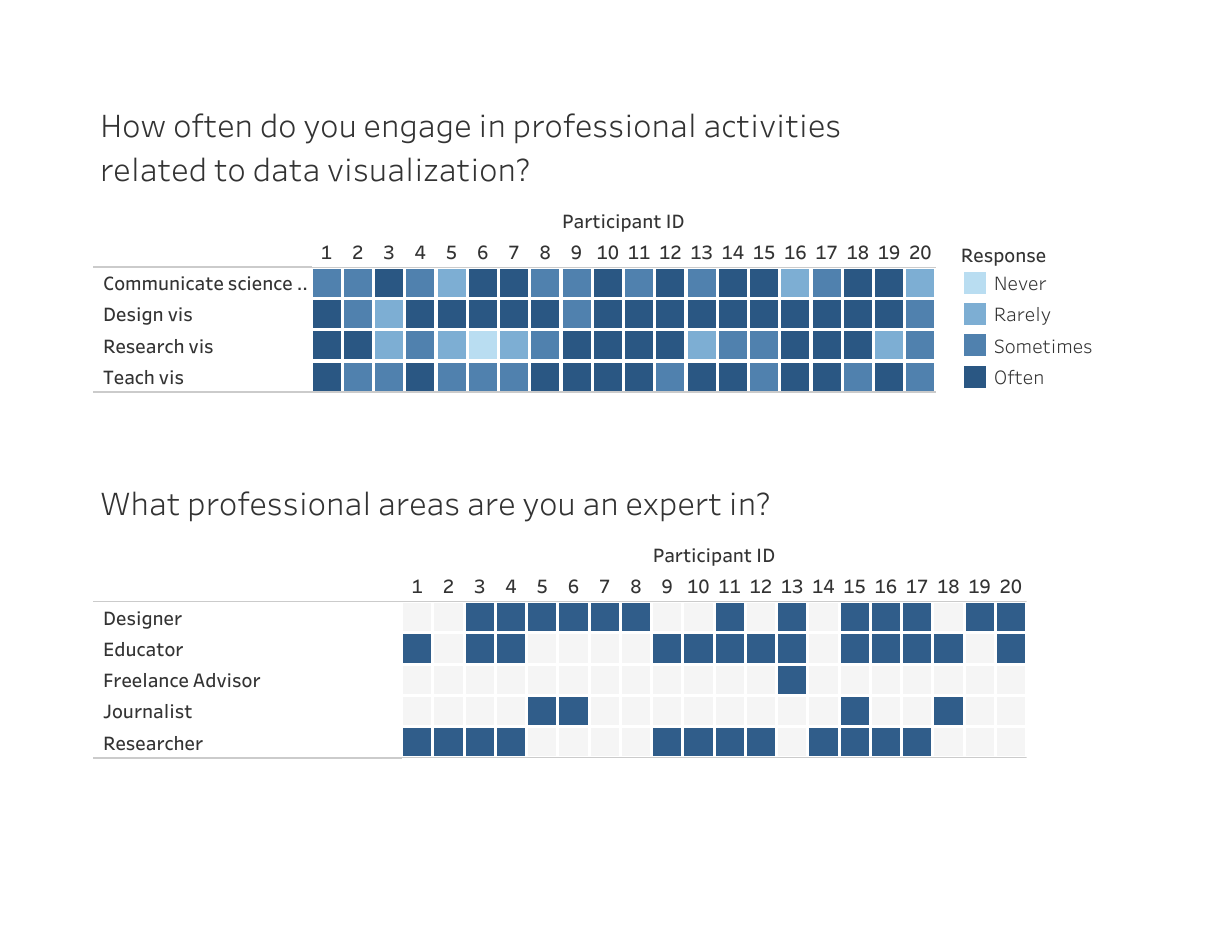}
    \caption{Demographic information shared by our interview participants.}
    \label{fig:demographics}
\end{figure}

Crucial to our task of identifying virtues was to speak with experts and other prominent \textit{exemplars} of ethical data visualization practice: ``we do not have criteria for goodness in advance of identifying the exemplars of goodness'' (Zagzebski~\cite{zagzebski2004divine} as cited in Hursthouse \& Pettigrove~\cite{sep-ethics-virtue}). As such, the three co-authors brainstormed a list of participants for our semi-structured interviews based on the following (qualitatively-judged) criteria:
\begin{enumerate}
    \item Ethical Focus: participants with prominent work in the space of data ethics and what it means to do good data visualization or, at a minimum, extensive experience with high stakes data visualization projects. Our final subject pool contains authors of books focused on ethical approaches to data visualization, academics who teach critical data visualization material in their courses, and consultants who produce guides and other industry-focused data ethics content, among other roles.
    \item Experience: participants with long periods of practice in the visualization field. In our final pool, the subjects had an average of 19 years of experience, and a minimum of eight years, in at least one focus area related to data visualization. 
    \item Diversity: We wished to speak with not just academic visualization researchers but also those likely to have fundamentally different conceptualizations of data visualization practice. Our subject pool included professors, data artists, data journalists, graphic designers, and consultants from three different continents.
\end{enumerate}

Our initial list was supplemented via snowball sampling---each interviewer was asked, off the record, who else should be interviewed. We recruited participants in batches via email; batching allowed us to be reactive to response patterns and, for instance, skew subsequent interviews towards practitioners if the previous round was heavily focused on academics.

Participants were asked in an initial survey to self-identify their area of professional expertise. We report their answers in \autoref{fig:demographics}. Given our narrow inclusion parameters and targeted research questions we did not collect additional demographic information such as age or gender; we discuss potential biases in our participant pool and interpretative approach in \S\ref{sec:limitations}.

\subsection{Interviews}
The interviews were conducted remotely via video conferencing and were recorded and transcribed with the consent of the participants. Each interview lasted ~60 minutes and was semi-structured. All of the interviews were led by the first author, with at least one other co-author present at all sessions to observe, take notes, and ask additional follow-up questions. The first two interviews, while included in our analysis, served as pilot interviews to inform the iteration of the interview questions. The interview questions were shaped by three key research questions:

\begin{enumerate}
    \item What values shape how data visualization exemplars think about their work?
    \item To what extent do exemplars experience ethical dilemmas between competing values? 
    \item To what extent do personal and professional values intersect? 
\end{enumerate}

These research questions, in turn, resulted in the following set of interview questions:

\begin{enumerate}
    \item In a very broad sense, what do you value in a data visualization; what do you think makes a ``good'' versus a ``bad'' graphic?
    \item In a very broad sense, what do you value in a data visualization designer? What do you think makes a ``good'' versus a “bad” designer?  
    \item What responsibilities do you think that data visualization designers have? Either to their audiences or to others?
    \item Do you consider the ethical or moral implications of data visualization work or making data visualizations? When in your workflow do you consider these implications? 
    \item What’s an example of a visualization or project you feel didn't meet your moral or ethical standards?
    \item What’s an example of a moral or ethical conflict or dilemma you had in your work? How did you resolve it?
    \item What are a few ethical considerations in visualization design work that you feel are overlooked or undervalued?
    \item Do you think your ethical values ever influence the visualizations that you make? 
\end{enumerate}

\subsection{Qualitative Coding}
Our qualitative codes were informed by an internal focus group with 16 visualization graduate students and researchers at our university, in which we asked participants to discuss a visualization they loved or hated, a visualization they felt conflicted about, and then had a general discussion of what they valued in visualizations (the materials and responses of this focus group are included in the supplemental materials). The following codes were discussed and defined collaboratively by all three researchers prior to beginning the coding process:

\begin{itemize}
    \item Clear values (single word): If the interviewee mentions an individual word that points to a value (e.g. ``effectiveness'')
    \item Pointing towards a value (phrase): If the interviewee talks about a value implicitly without necessarily mentioning a specific value word per se (e.g. they say ``I like when it doesn’t take very much time to read a chart,'' even if they don’t say ``efficiency'') 
    \item Value conflict: If the interviewee talks about a conflict, dilemma, or weighting between two sets of values (e.g. ``I like to make fun charts, but ultimately I have to decide if that’s worth it over making the chart easy to understand'')
\end{itemize}

Our coding procedure consisted of a dual-coded initial calibration phase with a subset of interview transcripts, followed by single-coding procedures for the bulk of the remaining interviews. Each author received two primary transcripts to code and two secondary transcripts to review. Upon completion of the first round of coding, we collaboratively reviewed the dual-coded interviews to discuss mismatches and disagreements. After this discussion, we felt confident in our coding structure and the ability to proceed with the full set of interviews with only single coders. Therefore, the remaining 14 transcripts were divided among the three authors again, but to a primary coder only.

We should note that our coding scheme and procedure was focused less on generating an exhaustive or fully mechanistic accounting of how people discuss ethics and personal or professional values, but rather on identifying the diversity of language and matters of concern we encountered, and providing points for mutual discussion and interpretation among the authors. As such, our codes and coding procedure are not strictly amenable to quantitative measures such as inter-rater reliability.

\subsubsection{Identifying Virtues and Values}

In all, 1,185 sections of our 20 transcripts matches to one of our existing codes, or were otherwise flagged for further discussion. Our goal was to condense these codes into a meaningful and diverse set of ``virtue words'' indicating matters of ethical concern in data visualization practice. This condensation into sets involved translation, interpretation, and discussion: while we would anticipate that many of the values we identified would be similarly identified by other analysts looking at the same data, we are mindful of the degree to which we bring our own positionality and interpretative lens to the transcript data.

The first author conducted an initial pass of value identification, based on the source text along with any additional comments or notes included as paratext by the coder(s) of the relevant transcript. In many cases, especially for items flagged with our ``clear values'' code, the translation step was straightforward. For example, this excerpt---\textit{``do people understand what they need to understand out of it?''}(P8) clearly pointed towards the existing value ``Understandable.'' P13, likewise, explicitly uses a list of virtues in their teaching and consulting work, which they shared with us during our interview, and which was readily incorporated into our existing list. In other cases, the values were implicit. For example, P7 discussed the idea that before making a visualization \textit{``we should first figure out if the data supports what you have in mind.''} While this quote doesn't mention the word ``honesty'' explicitly, this participant is pointing towards the idea that the charts we make should honestly reflect what the data supports. The three authors met to discuss cases where converting an excerpt to a value was not clear and resolve to a final code. In addition to values, authors also kept track of emergent themes and recurring patterns of ethical practice.

We generated a final list of 68 matters of concern, each of which we describe as either applying to an object (one might value a visualization that is accessible, for instance), to the designer (one might value a designer who is humble), or both. We then conducted affinity diagramming, clustering values together according to semantic and conceptual similarity. Through a series of discussions, we iterated over value placement until all three authors agreed on the placement of the values into a final set of nine clusters. For the remainder of this paper, we refer specifically to those 68 lower-level matters of concerns as \textit{values} (as a term broader than ``virtue'', and to avoid having to make the normative judgment of which are ``genuine'' virtues, and which are merely statements of importance), and the nine clusters as \textit{virtue clusters}. We organize our discussion in the following section around these virtue clusters. \autoref{fig:table} enumerates our values and virtues. Specific definitions of each value are included in the supplemental material.

\section{Results and Virtue Clusters}
    
\begin{figure*}[!ht]
    \centering
    \includegraphics[width=\textwidth]{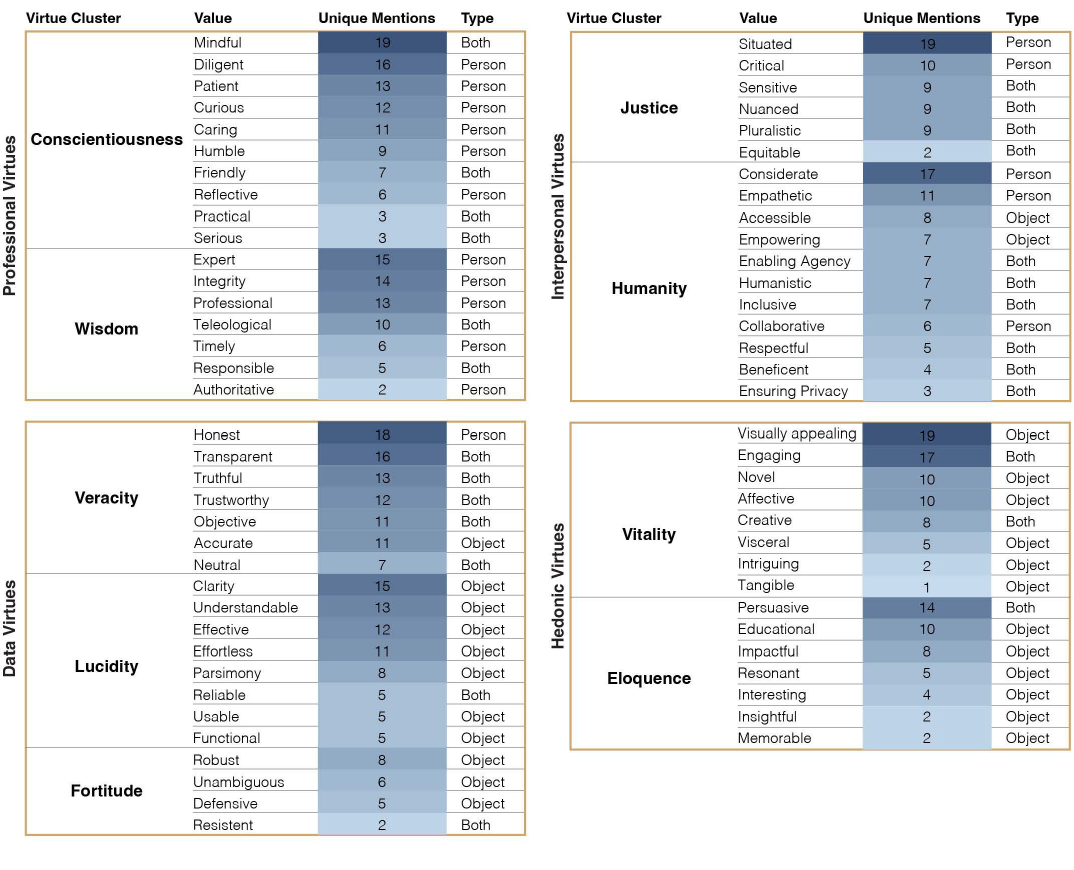}
    \caption{A table indicating values arranged within semantically similar virtue clusters. We also indicate how many interview transcripts (out of 20) contained codes associated with each value, and whether this value applies to the specific visualization \textit{object} being created, the \textit{person} designing said visualization, or could have a meaning that applies to \textit{both} designer and visualization. Consult our supplemental material for our short working definitions of each virtue.}
    \label{fig:table}
\end{figure*}



In this section we present themes that emerged from our interviews and recurring considerations of virtues of importance to visualization work. We omit a full accounting of each of the 68 values for space and reader attention, although our working definitions of each are included in the supplemental material. Instead, we use virtue clusters of semantically similar values to be our primary unit of discussion. In the following four sections, we discuss sets of virtues clusters organized into groups denoting different areas of ethical concern. Within these sections, we provide a brief description of each virtue cluster and their ethical underpinnings. We then discuss emergent themes that recurred across our interviews that deal with the relevant values within these clusters.

It is important to note that these clusters are \textit{affinities}, not strict or exclusive ontologies: many of these concepts are necessarily interrelated, mutually enforcing, and display a kind of recursive relationship with each other. For example, situating data in context is a way to be more transparent, which builds trust and supports communication---and if you want to be more trustworthy, you might provide contextual information about data.


\subsection{Professional Virtues}
\label{sec:wisdom}

The virtues in this group, wisdom and conscientiousness, draw upon Aristotle's definition of phronesis, or practical wisdom. In \textit{Nicomachaen Ethics}~\cite{aristotle_nicomachean}, phronesis is a foundational virtue that guides the application of other moral virtues. Phronesis can be thought of as the wisdom that guides one to ``deliberate well about what is good''~\cite{wisdom_2023}. Gage~\cite{duffy_virtue_2018} writes that phronesis ``is the meta-virtue without which other virtues may not be able to be acquired or practiced''. Peterson and Seligman~\cite{peterson_character_2004} consider wisdom to be a core virtue, defining it as ``knowledge hard fought for, and then used for good.'' Values in these virtues clusters includes dispositions and skills that must be built up over time and through experience as well as broader, practical attitudes towards visualization design.

\subsubsection{Building Self-Awareness \& Maintaining Professionalism}
Wisdom is related to being able to understand your subject and data and having a strong awareness of your skills and limitations as a visualization designer. A designer should have the capacity to understand the data, or to talk with people who do and get into conversations with domain experts, or, at the very least, the intellectual humility to understand their own gaps. P7 remarked that they like to learn as much as they can \textit{``about what the data represent, and if that dataset can actually say something meaningful about the topic,''} while P5 mentioned that, when hiring, they see it as a strength for a designer to be \textit{``familiar with the language of the area that the data is originating from.''} and likewise P19 emphasized that we should be \textit{``taking the time to understand how data was collected''}. This awareness can also impact what projects designers take on, as P14 reflected, \textit{``I don't want to say to a potential client that I can do this work if I don't feel like I can do it at the highest level,''} and so there is an aspect of knowing \textit{``your own limitations in your knowledge, your skills, your time.''} 

Navigating personal connections to one's work was discussed. Some participants emphasized maintaining some separation---you need to understand as a designer \textit{``that your idea is not a testament of your own personal value''} (P20), and \textit{``I try to take the critique or the criticism not personally''} (P16). Others emphasized that some aspect of personal connection is unavoidable---\textit{``the convictions you have as a designer always shape what you do, and I think it's not something you can even turn on or off, so it just happens''} (P7). One interviewee argued that it is possible to land somewhere in between: \textit{``you have to sort of separate yourself from that a little bit, but at the same time, that's what makes a good designer, is having that personal connection''} (P3). 

Conscientiousness is related to the general attitude one approaches the process of visualization with, for example, an attitude of curiosity and care towards data and the entire process of designing with data is ethically valuable, per P10 \textit{``A good designer is someone who cares and pays attention and is curious.''} Visualization work is often completed under a tight deadline, which can negatively impact the conscientiousness with which a designer might normally operate. P5 reflected that \textit{``those things that are just little inklings of, this doesn't feel quite right. Those might get squelched if they happen too close to the deadline schedule,''} while P6 stressed the impact of deadline pressure, especially when you have a boss \textit{``leaning over you saying, where is it? You know, we need it. We needed it yesterday.''}
Accounting for crunch ahead of time is one way of maintaining professionalism: P20 noted that in some cases, a designer \textit{``might just be tired, and busy, and too many things at the same time, and errors always trickle in. So finding ways of testing and validating the design before it actually gets out into the world normally goes a long way.''} 

\subsubsection{Saying ``No''}
Professional integrity involves navigating refusal: to able to push back, navigate ethical conflicts, and know that they may face potential ramifications for doing so. P17 stated that \textit{``the practice of saying `no' is really important.''} P6, in describing a past project where they were asked to truncate the y-axis, concluded that they\textit{``had to say no, because sooner or later, somebody's going to say, who did this?''}, and later asked \textit{``at what point do you withdraw if you have this sinking feeling that this isn't right?''} P8 similarly reflected that designers have to ask themselves, \textit{``is this whole project in the first place the right thing to do?''}  This refusal can even be preemptive: P7 advised that designers should ``be super strict at the beginning of the project about who you collaborate with and in which context''; as a warning about not doing so, P7 related an experience where they had misgivings about a project but ``\textit{I was not in a position any more to pull back.}''

Navigating the pressures of the job and client expectations in connection to ethical considerations was further discussed by a number of participants. P8 described a client who expected that they \textit{``purely abide by their point of view, not challenge them,''} stating that they respected their opinion, but disagreed with it. P18 speculated that \textit{``somebody with power is going to push you to do something you don't want to do. You're going to lose your job if you don't do it? Maybe. Can you live with yourself if you do it?''} P20 emphasized that \textit{``you have to be smart about that dance, you have to be picking your battles, you have to go around, show them otherwise. Maybe that this is not the right theme, this is not the right topic, and here's another solution, why don't we do this?''}; P7 likewise suggested \textit{``if you offer a problem without a solution, you're not gonna get anywhere.''}
Beyond just refusing to take jobs or compromise design principles, there was also discussion of when personal responsibility involves active acts of resistance. For instance, P12 said \textit{``I don't know people on the inside of the government who know where ICE is going to go next. But if I had that information, and I was working for the government, would I view it as my moral responsibility to feed it to people who could alert people? 100 percent.''} 

Participants, however, were cognizant that ethical principles of refusal can be in conflict with pragmatic considerations. P14 said, \textit{``I would not tell people that they should turn down work that they need to put food on the table,''} and that if you take a job that does not align with your personal values, you can try to focus on doing it as honestly and objectively as possible. P20 reflected that you have to set your own boundaries, based on your principles and moral limits: \textit{``you're going to sleep with yourself every night, right? And you're going to know if you've done the right thing or not. That's for me what matters.''} While the virtues and values discussed in this work represent ideals that practitioners aspire to, these ideals are often tested by practical realities, from client demands, to power imbalances and economic pressures. The tension between the ideal and the actual~\cite{mills_ideal_2005} can constrain one's ability to act in an ethically ideal way, regardless of a designer's personal convictions. 



\subsection{Data Virtues}
\label{sec:datavirtues}
The values within these virtue clusters are perhaps the most familiar to the visualization research community, building off common concepts like Mackinlay's~\cite{mackinlay1986automating} notion of effectiveness and expressiveness, the admonition to clearly present data~\cite{kostelnick_visual_2008, tufte1983visual}, and the desire to avoid misleading or deceiving with ones charts~\cite{lisnic_misleading_2023,correll2017black}. Of note are the last group of values in the Fortitude virtue cluster, which involve consideration of how a visualization may be interpreted when it is removed from its original context or initial conditions (for instance, a visualization that is ``defensively''~\cite{correll_teru_2023} constructed to not misrepresent the data across artifacts of sampling error or design parameters).


\subsubsection{Making Things Clear}
Interviewees consistently emphasized that readers should not struggle to understand visualizations, or, worse yet, should not misinterpret visualizations. P5, P9, and P19 noted that good data visualizations are clear, make something clearer, and help with understanding. P18 highlighted that data visualizations can reveal \textit{``something that people may not otherwise understand.''}.
Beyond just understanding, there was also value placed on understanding data \textit{quickly}. P15 underscored that \textit{``you need that quick read, you need people to understand,''} while P5 warned that \textit{``the minute somebody is kind of like, `ugh, this is hard,' we're probably going to lose their attention.''}  P19 framed this as hitting the right balance of \textit{``speed to insight.''} Yet, participants resisted treating speed as the only measure of effectiveness. P20 introduced a counterpoint: if something takes longer to learn, it is more likely to be remembered, meaning \textit{``there's actually an advantage of having an effort to understand a graph''} in some contexts.

Participants also noted concerns around audience and data literacy when considering the effectiveness of a visualization. P17, for instance, stated that (\textit{``when someone shows an animated scatterplot in a government public meeting, that's usually a failure, for my perception, because most people in the room can't understand it''}), while P11 asserted that designers bear a certain obligation not just to present information, but to \textit{``take [people's] hand and walk with them. If we are just waving from the other side of the cliff, then we're also not being very effective.''} 

Aesthetics also entered this conversation. Some participants treated beauty as separate from, or at odds with, effectiveness (\textit{``You can make the most beautiful dashboard in the world, but if the person or people you want to use it don't actually go do it, then it's not an effective visual''} (P14); \textit{``It looked great, it was very creative. But is it really something that would serve the business leader at the company to make choices? I'm not convinced.''} (P8)). P20, however, pushed back on this framing, arguing that \textit{``people think about aesthetics versus usability, form versus function, when it's not. They are not two ends of a spectrum.''} 

\subsubsection{Striving for Active Honesty}
\label{sec:honesty}
A duty towards honesty was paramount. P6 argued \textit{``that's the main thing actually. Tell the truth,''} while P18 emphasized that one should always focus on presenting facts, P17 stated that data visualization should be a \textit{``projection of knowledge and correctness.''}, and P2 asserted that at its core, a visualization has \textit{``got to be right...reliable and accurate.''} Designers were cautioned not to distort the data (P14), avoid deliberately misleading tricks (P19), such as cherry-picking data (P5) or truncating the y-axis, and generally avoid things that could result in people believing incorrect information (P7). These concerns echo Tufte's foundational work on graphical integrity, which similarly cautioned against distortions that give readers a false impression of data~\cite{tufte1983visual}. However, for P3, this drive towards honesty involved more active searching for truth and insight, rather than a mere passive duty to avoid deception:
\textit{``there's a responsibility to the data to give it the best chance to reveal itself and its patterns.''} However, the desire to represent data with fidelity can be counterbalanced by cognizance of its limitations. Per P3, \textit{``I feel like the responsibility is to the data. But the data is not neutral anyway.''}

One observed exception to perceived duties related to truth and honesty was uncertainty information. P10 reflected that visualizing uncertainty can be particularly complicated, and that showing too much could lead a viewer to \textit{``completely disengage from the whole thing.''} 

An interesting recurring metaphor that suggests that truthfully presenting the data is not a rote matter of encoding data values was the notion of ``smells'' (as in ``data smells''~\cite{foidl2022smells} or ``information scents''~\cite{pirolli_effect_2000}).
For instance, P5 described misgivings with a project in nasal terms: \textit{``This does not smell right. Something weird is going on''}. P13 highlighted a similar concern related to the importance of a designer being willing to say \textit{``something's off here, I'm going to go dig.''}~\cite{pirolli_effect_2000} Honesty in this sense may involve acquired instincts born of long experience: a matter of phronesis rather than a simple avoidance of deception.

\subsubsection{Building Trust}
Transparency emerged as a foundational aspect of building credibility and trust. As a design feature, participants described transparency as an active practice: being \textit{``clear about where the data has come from and what is and is not represented,''} (P2), deploying annotations, footnotes, and other mechanisms to help readers better understand the data (P2, P5, P9), and ensuring viewers can trace the underlying data and understand which facets were selected (P12). Several participants framed transparency not just as good practice but as an ethical obligation. P7 described it as \textit{``not lying by omission,''} that is, being transparent that there is always some kind of authorship happening in visualization. This kind of transparency was seen as directly building credibility, per P18: \textit{``having a source, a clear source, and a clear methodology, and clear labeling, and transparency is 100\% your credibility...it's important ethically to be clear about what your source is and to be transparent about what you're showing.''} P5 and P9 emphasized the importance of resisting the temptation to present data as cleaner or more perfect than it is. To combat this, they recommend noting missing categories, flagging unreliable values, and being forthright about what a dataset does not capture. 

\subsubsection{(Re)considering Objectivity}
\label{sec:objectivity}
Aiming for impartiality and objectivity was another recurring pattern, and in particular an admonition to be \textit{``true to the facts''} (P20) by faithfully presenting the data. P18 argued that \textit{``data is not something that you can force into a box---it will tell you what it is,''} while P8 observed that \textit{``at some point, the data says what the data says.''} A lack of room for subjectivity was occasionally seen as a goal: per P13, \textit{``I aspire to disappear.''} Likewise, for P10, \textit{``If you've done it well, then you've made it so clear that it's obviously objective.''} 

Yet, there was also recognition of the impossibility (and in some cases non-desirability) of total objectivity. P17 stated that they are against \textit{``the idea that a dataset is some objective picture of reality''} and P9 went further, describing data not as truth but as \textit{``a produced artifact in the world, produced through decisions and people and instruments and history and power.''} P3 challenged the assumption that data speaks for itself: \textit{``people think that if you just put the right data in the right chart, it will speak for itself, but I don't think that's true at all.''} P7 warned that pretending the designer does not matter \textit{``because everything is data and effective''} leads into \textit{``muddy and dangerous waters.''} P13 noted that despite the field's tendency towards neutrality, emotions are always present---but that\textit{``some people feel them more strongly than others.''} Other respondents who carved out room for explicit advocacy and partiality: per P12, \textit{``We're trying to make an argument, so we don't necessarily want to be as objective as possible''} and, likewise, \textit{``visualization is rhetorical. It can be more or less rhetorical, given the context, but it's never not rhetorical.''}
P17 drew a careful distinction between perspective, which may be unavoidable and legitimately inform practice, and bias, whereas P13 stated: \textit{``I operate from a certain framework...I'm still biased, and I know it.''} For some, this tension resolved not into an answer but rather into a kind of humility: P9 reflected that \textit{``the more I learn about different perspectives and different ways of framing the problem, the more it's clear to me there are no right answers. The best I can do is just acknowledge and be aware and try to make the best decisions I can.''} Together, these responses suggest that objectivity functions less as a fully achievable state and more as an orienting value: something that participants strive towards while remaining cognizant of limitations.

\subsection{Interpersonal Virtues}
\label{sec:interpersonal}

While the values within our Wisdom and Conscientiousness virtue clusters involve personal and professional betterment, the clusters in this group involve \textit{interpersonal} considerations. 
Justice and Humanity both involve considering and improving the well-being of others. Justice involves broadly considering fairness and equity within structures systems, whereas humanity is more directly oriented towards interpersonal relationships \cite{peterson_character_2004}. For instance, a justice-based concern with a data visualization might be if a \textit{group} is harmed or stereotyped by a visualization, whereas a humanity-based concern may focus on an \textit{individual's} ability to benefit from a visualization. A concept connected to these interpersonal virtue clusters is building an \textit{ethics of care}, an often-overlooked component of visualization research~\cite{akbaba2023troubling}, as well as the principles of Data Feminism~\cite{dignazio_introduction_nodate}, in particular its emphasis on examining power, embracing pluralism, and considering context, which resonate closely with the Situated, Pluralistic, and Equitable values in the Justice cluster, and its emphasis on elevating emotion and embodiment, which aligns with the Empathetic and Humanistic values in the Humanity cluster. 

\subsubsection{Situating the Data}
\label{sec:situated}
Participants emphasized the importance of always situating and understanding data and visualizations contextually. P3 simply stated that \textit{``context is huge,''} while P17 observed that \textit{``visualizations always exist within a larger context''} and that \textit{``everything's connected to that context.''} P10 identified the absence of contextual information (sourcing, framing, auxiliary elements) as a defining feature of bad visualizations, and P18 underscored that facts are never simply facts: \textit{``They're the facts within the context of other facts.''} Several participants described context encompassing not just the data and its topic, but also the audience (P17), the community being designed for (P9), and the broader situation in which a visualization will be encountered (P12). 

For many participants, taking a situated perspective meant asking critical questions about data provenance. P12, P15, and P9 pointed to questions of origin and power—where the data came from, who funded its collection, who has authority over what counts as data. P9 further argued that designers have \textit{``a responsibility to make people aware that data is imperfect and partial,''} and P14 and P15 both emphasized the importance of taking a critical eye towards data and being willing to actively investigate rather than accept data at face value. 

\subsubsection{Centering People}
Participants expressed a strong sense of responsibility towards both the people represented by data, as well as the people in their audience. P17 says they ask themselves \textit{``how would I feel if I was represented like this, if I was in this data set?''}, echoing guidance in the literature \cite{schwabish_no_nodate}. P5 noted that they are \textit{``very attuned to this idea of data points representing humans,''} and P12 emphasized that they \textit{``think really hard about the encoding of data points relating to people.''} Moving beyond how people are encoded by data and representations, P14 proposed designing in ways that enable people and communities to feel \textit{``valued and respected,''} and P8 raised the concern of inadvertently reproducing stereotypes through encoding choices, such as using pink and blue for binary gender. P5 stressed the importance of protecting individuals' privacy, P19 described the practice of ethically removing subpopulations that could become identifiable, and P10 identified a failure of anonymization as a clear ethical violation. 


Beyond concern for the people behind the data, there was also broad recognition of the need for concern for one's audiences. P19 asserted the value of \textit{``meeting people where they are, with the information they need when they need it,''} and argued that making good charts so that people can make good decisions is a core obligation. P13 framed it as a question of rights: \textit{``What rights does a person have when they read this visualization, and what tools do they need to truly understand it. Are we truly providing it?''}
P15 reflected that they ask themselves: \textit{``Am I meeting the audience where they're at and not flying over their heads?''}, and P9 urged designers to avoid making people feel stupid. Several participants described their sense of responsibility towards their audience in expansive and relational terms. P8 spoke about building a shared language and understanding what audiences need without passing judgments, and P17 described making things that people \textit{``can gather around to work with and reflect on and digest.''} 

Participants also acknowledged limitations. P8 noted that \textit{``there's just no true perfect accessibility,''} and designers often make sacrifices based on project constraints. P17 voiced concern for understanding how different audiences might understand or relate to the same visualization. These considerations connect to concerns about data literacy~\cite{creamer_finding_2024}, accessibility \cite{joyner_visualization_2022}, and reflect the call for a wider definition of what makes a visualization effective, based on ``how, by whom, where and for what purpose visualizations are encountered''~\cite{kennedy_engaging_2016}.

\subsection{Hedonic Virtues}
In these virtue clusters, we focus on values associated with hedonics, aesthetics, and the other parts of visualization design above and beyond the staid communication of data values. Per Aristotle, the logical content of an argument is only one form of persuasion: \textit{pathos}, or the appeal to emotion, is another important consideration. Prantl et al.~\cite{prantl2026untangling} in particular claim that the scope and function of pathos is underexplored in visualization research. Even without specific consideration of aesthetics or emotional components of a visualization, part of being a good data \textit{communicator} may involve being a good data \textit{storyteller}.

\subsubsection{Considering Rhetoric}
\label{sec:persuasive}
The persuasive potential of visualizations, and so the power of visualization designers, recurred as a theme: per P13, \textit{``whoever makes the vis[ualization] owns the message,''} and the visual elements included should therefore support that message. P17 extended this idea, arguing that because every dataset contains multiple possible stories, the choices about which story to tell and how are precisely where visualization overlaps with ethics. For P11, rhetoric is related to how visualization \textit{``can be reflective, it can be interpretive, it can be responding to the idea of the world that it is trying to represent. It is not made of zeros and ones.''} P11, when describing the construction of a data story, even mentions ``\textit{manipulating}'' the reader by intentionally suppressing information until later in the story, in service of ``\textit{a deeper truth, an emotional truth.''}


Ethical rhetorical practices were described as a process of honest investigation and argument. P17 emphasized diving into the data to explore the potential stories in it, P14 described being truthful in itself as a potential form of persuasion, and P18 proposed that finding stories in data requires \textit{``understanding where it fits into the context of what you want to impart to readers.''} P13 emphasized alignment between the visual representation and the message, rather than speed of comprehension (\textit{``I don't care if it takes 1 second or 10 seconds to understand it, I care more about whether the systems that you use are supporting the message you intend to display.''}) P10 likewise warned that dogmatic adherence to perceptually optimal forms like bar charts risk producing \textit{``a really one-dimensional story.''}

\subsubsection{Navigating Novelty and Engagement}
\label{sec:engaging}
A recurring tension across participants' responses was the perceived divide between serious, functional visualization and work that is engaging, creative, or affective. Several participants described adjusting their standards based on where a project exists on this spectrum. P2 distinguished between a visualization \textit{``thrown on Tableau Public to look cool''} and one built to support a hospital team, suggesting that the seriousness of an application shapes how carefully it must be executed. P5 framed it as a question of purpose---whether a visualization is meant to be useful or to inspire \textit{``curiosity, awe, and wonder,''} while P8 treated creative elements as essentially decorative, describing them as \textit{``fluff.''} P16 reflected that \textit{``we're not always trying to push the boundaries. Sometimes we just want to make it right,''} indicating that these are separate commitments. P9 likewise distinguished those visualizations made from following standard practices and more expressive or engaging visualization work: (\textit{``following the rules and structures, you're not going to get a glorious `Giorgia-Lupi[-style]' visualization. You're not going to get that.''})

Other participants challenged the underlying assumption that these values are in opposition. P20 argued that \textit{``beauty is a central element of functionality,''} while P11 noted that the ``exhibitionary'' mode (often seen as the ``bad kid'' out of three approaches to research and analysis) has \textit{``been proven again and again as being so essential for effectiveness.''} P9 described the goal of getting people to \textit{``slow down and engage''} as a serious aim in visualization. The ongoing negotiation across this binary can be characterized by P14's pragmatic comment, noting that we always need to \textit{``balance the truth and the data with trying to engage people.''} P8 likewise describes an intentional choice to sacrifice detail and accuracy in order to prioritize messaging, understanding, and the emotion of hope, and then later emphasized that aesthetic elements are essential layers to add to accurate charts ``\textit{to make sure that people don't run away from the data.}''

\section{Discussion}




Despite our repeated caveats around the limitations of our list generation process and the impossibility of truly exhausting all of the matters of care that could potentially arise in the process of doing visualization work, we nevertheless generated a diverse list of virtues capturing a multifaceted set of concerns across a number of critical roles in our community. From the data artist seeking to develop a new and engaging piece for an exhibit, to the scientist seeking to accurately and honestly display experimental results, to the data journalist working under time pressure to get out an article when and where it can do the most good: all might strive to embody some or all of the virtues our participants discussed. Having such lists can provide pedogogical value~\cite{newstead2020evaluating} and work as a both a tool for expanding ethical considerations when conducting a technical project~\cite{conwill2025design} and for brainstorming ethical challenges for emerging data science work. For instance, in work that inspires the tarot-card based format of our teaser figure (\autoref{fig:teaser}), Wang et al.~\cite{wang2024card} employ a card-based approach to investigating ethics in AI for data visualization, with a set of 11 ``principle cards'' that stochastically surface ethical matters of concern. While we are under no illusion that any member of the visualization community, no matter how ethically-minded, is consciously weighing their work across dozens of ethical virtues simultaneously, we maintain that this stochastic surfacing of principles can be useful.

As interesting to us as the \textit{diversity} in values were the observed \textit{commonalities}: the values that recurred across nearly all of our interviews. Nearly all (N=19) interviewees, for instance, discussed (in some form or another) being \textit{Mindful} of the biases inherent in data, and the \textit{Situated} nature of knowledge, suggesting to us that admonitions from the critical data science community~\cite{correll_ethical_2019,d20206,gitelman2013raw} to challenge the perceived neutrality and objectivity of data have borne fruit (or, just as possibly, that such illusions about the objectivity and perfection of data do not survive multiple years of experience in the visualization field). A concern for aesthetics, and making \textit{Visually Appealing} visualizations was equally common (N=19 interviewees), likewise suggesting that, beyond a concern (only) for effectiveness or efficiency of data visualization, the ability for a visualization to evoke the right emotions and messages through aesthetics was recognized as another core goal. Despite this centrality, the aesthetic~\cite{carlsson_aesthetic_2010,cawthon_effect_2007,kostelnick2020art, lan_more_2025}, emotional~\cite{prantl2026untangling, kennedy_feeling_2018, campbell_feeling_2019, lan_affective_2024, lee-robbins_affective_2022} and empathetic~\cite{boy2017showing,morais_showing_2022,zer2015dataviz} potential of data visualizations are understudied in the academic visualization community.

The framework we present is not prescriptive in the way that other approaches to ethics, such as rule-based deontological approaches, aim to be. Merely listing, for instance, that being ``transparent'' is a relevant virtue for data visualization does not provide sufficient guidance to determine when or if, for instance, to incorporate uncertainty information in a chart. We do not wish to tell designers of data visualizations what to do. This is a deliberate choice that is consistent with our argument that ethical visualization is a matter of judgment informed by building practical wisdom and expertise, but it does mean that the framework will not provide easy answers for readers seeking rule-based guidance. However, making these virtues and values explicit can support designers and members of the visualization community to reason about them deliberately rather than as implicit intuitions. These virtues can also be used to guide actions and decision-making~\cite{swanton_virtue_2001}: we can, for instance, look for the personal virtues present in a visualization designer we admire and seek to emulate them, or weigh a choice between two visualization designs based on which better embodies the object virtues we wish to prioritize in our work. We also highlight that some of the virtues discussed in \ref{sec:datavirtues} and \ref{sec:interpersonal}, particularly around being transparent, considerate, and ensuring privacy, overlap closely with the principles of the UK Government's Data and AI Ethics Framework~\cite{data-ai-ethics-ukgov}. Notably, this framework also emphasizes that the principles can come into tension with one another, requiring contextual judgment and trade-offs rather than fixed rules. 

As a last point, we highlight the inescapability and ubiquity of personal values in visualization work: values are not merely abstract concepts but directly impact the kind of visualization work that is performed in the real world. We asked each interviewee if they thought their personal values ever shaped their work in visualization: most of the participants did not just acknowledge that this is true, they went so far as to insist emphatically upon it: \textit{``100\%''} (P8, P11), \textit{``almost always''} (P12), \textit{``all the time''} (P15, P17), \textit{``totally''} (P13), \textit{``absolutely''} (P7), \textit{``of course''} (P16). When we asked the question, many participants first laughed as though the answer to the question was obvious, and P11 replied, \textit{``did anybody respond no? Can you respond no?''} Two participants expressed explicit hope that their values had an impact on their work, saying \textit{``I hope so''} (P10, P11). In short, the view was consistent: our visualization work is shaped, implicitly or explicitly, by the values we bring, and the people we aspire to be. Just as AI systems reflect the values of the societies that create them~\cite{lee_mirror_2026}, our values may also be reflected in the visualization artifacts we make. 



\subsection{Tensions}

In addition to revealing a series of values and virtues in visualization, our exploration also exposes salient tensions that visualization experts often encounter and must navigate, often with little formal guidance. The interdependence of the values makes them difficult to resolve, because pulling on one value necessarily effects another. We do not seek to resolve these tensions; they are constitutive of the practice. Therefore, rather than viewing these tensions as failures, we present them as opportunities to reflect upon the complex balancing act that visualization designers negotiate in their work. Through highlighting these tensions, we surface gaps and discuss opportunities for developing further ethical guidance in pedagogy and practice. In participants' own words, there are \textit{``no right answers,''} rather it is about \textit{``knowing enough to see the tensions''} (P9). These things are all \textit{``part of a spectrum... levers you can adjust''} (P18). If ethical visualization is less about applying rules and more about building the judgment necessary to navigate this kind of complexity with awareness and care, then cultivating ethical action necessitates building the conditions in which this judgment can develop in pedagogy and practice. 


\subsubsection{Subjectivity versus Objectivity}
As discussed in \S\ref{sec:objectivity} and \S\ref{sec:situated}, we observed a tension between striving for objectivity and neutrality in one's visualization work versus a recognition of the inherent biases and positionality of data collection and visualization. The desire to achieve full objectivity runs deep in Western culture \cite{haraway_situated_1988}, and visualization is no exception---objectivity is often an implicit expectation \cite{saharan2026critical, zhang_deconstructing_2025, kennedy_work_2016}. Yet, our participants almost unanimously mentioned values that problematize this apparent objectivity, while at the same time also stating duties they had towards avoiding a perception of partiality or bias. Our participants condemned both extremes: that of coming into a data visualization project with a preconceived story, and twisting the data to fit that story (this specific practice was a reason why at least two of our interviewees backed out of consulting jobs), but at the same time critiquing data visualization designers who uncritically assumed the objectivity of the given data.

\subsubsection{Data versus Truth}
A related tension as discussed in \S\ref{sec:honesty} was how to navigate the (often strongly felt and unambiguously stated) duties to faithfully represent the data, while at the same time being honest or truthful in ways that involve disclosing the limitations, biases, and uncertainty in these very same data. The observation by Hullman~\cite{hullman_why_2020} that visualization designers regularly exclude uncertainty information for reasons of reducing complexity, potential misinterpretation, or to better reach their perceived audiences is a concrete example of this tension in action: is the omission of this data dishonest, or a practical necessity if the important message \textit{in} the data is to be successfully communicated? A connected example were the interviewees who praised those visualization who dug deeper, who, in keeping with the analogy of ``data smells''~\cite{sultanum2024data}, did not take the data at face value but dug deeper (perhaps dispelling any superficial ``visualization mirages''~\cite{mcnutt_surfacing_2020}) to uncover latent patterns or deficiencies. Visualization work, then, must be true to the data but at the same time not \textit{beholden} to the data as the only source of truth.

\subsubsection{Persuasion versus Neutrality}
Disagreements or conflicts around both honesty and objectivity were also felt when considering the role of rhetoric and persuasion in data visualization. As discussed in \S\ref{sec:persuasive}, there were those who wished to ``disappear'' in the work and present on what were seen as the inherent patterns in the data, and a discomfort with designers who stray too far in advocating for a particular view (especially prior to the collection of any data). Yet others took intentional steps to increase the impact and persuasive power of their work, shaping intentional data narratives meant to guide the reader along particular paths. Ethical and unethical (or at the very least virtuous and vicious) behavior could conceivably exist at both extremes. Adding more potential complexity to this tension is what Kennedy et al.~ \cite{kennedy_work_2016} call the rhetorical ``work'' of data visualizations, where potentially messy or controversial data are reduced to clean lines and marks on a page, seemingly presenting little room for disagreement, counterargument, or critique. Puerta et al.~\cite{puerta2025many}, similarly focus on the ``implicit rhetorical work'' performed by charts to suggest that there is not a hard and fast distinction between overtly ``propagandized'' visualizations, and seemingly ``neutral'' ones.

\subsubsection{Engagement versus Efficiency}
Lastly, there were tensions, both within and between interviews, on the role of form versus function, or whether making a \textit{useful} visualization was or was not at odds with making a \textit{beautiful} one. An interesting recurring pattern was the degree to which these decisions seemed to come down not just to personal taste or intended audience but the perceived \textit{genre} of the visualization. In keeping with Gelman \& Unwin's~\cite{gelman2013infovis} notion that infographics are ``the Armani suits'' to statistical graphics' ``clunky plaid shirts with pocket protectors'', people who were willing to explore artistic forms and methods of engagement in data journalism and data art were unwilling to do so in other contexts such as business intelligence dashboarding or science communication. We note that the presumed default of minimalism in data visualization is not an inherent, neutral, or even empirically unambigious good~\cite{hill2018minimalism,inbar2007minimalism,parsons2020data}. However, the idea that engagement and efficiency are inherently opposed seems widespread (although certainly not universal) among our interviewees.

\subsection{Limitations and Future Work}
\label{sec:limitations}

One limitation of this work is the inherent incompleteness and subjectivity of our process of generation and analysis. This particular set of concerns emerged from our sample of participants. Other communities of visualization practitioners, educators, and researchers, or studies using different methods, could surface different or additional values. We reiterate again that our research goal was not to generate a comprehensive and authoritative list, but merely a generative and ecologically valid one. We note, too, our own positionality in this work, which by necessity limits the breadth and depth of our ethical analyses: we come to this project with the perspective of Western data visualization researchers from information design and computer science backgrounds: given how entangled values and virtues are with one's social, political, and religious milieux, we are hesitant to overgeneralize.

While our interview study generates a set of matters of concern, it does not map in detail the extent to which considering these values and tensions could translate into guiding decision making during the visualization design process, or why some values necessarily are weighted more heavily than others in specific contexts. Further empiricism is needed. For instance, observational studies---following designers through real projects rather than asking them to reflect retrospectively---could deepen this understanding significantly. That is, what people (even experts) \textit{say} they value is one thing: it remains to be seen to what extent these values are embodied or prioritized \textit{in practice}.

Lastly, we wish not only to \textit{enumerate} values, but to place ethics centrally in the \textit{practice} and {pedagogy} of visualization. Per Marx~\cite{marx2024theses}, ``Philosophers have hitherto only \textit{interpreted} the world in various ways; the point is to \textit{change} it.''
While we find the framework of values generative in our own practice, giving us structured language for exploring and articulating ethical considerations (e.g., ``which values do I emphasize more heavily, which ones am I less concerned about, and why do I weigh those values in this manner?''), we are particularly interested in how virtue-based frameworks could be applied to data visualization pedagogy, although translating our framework into a structured pedagogical resource is beyond the scope of the current work.

\subsection{Conclusion}

The field of data visualization has made progress in identifying (and avoiding) misleading or deceptive visualizations. However, the focus on avoiding overtly harmful acts captures a limited view of the ethical landscape that visualization experts actually navigate. We address this gap by surfacing a rich set of values through 20 interviews with visualization designers, artists, researchers, educators, and journalists. The result is a framework of values and virtue clusters that reflects matters of concern that visualization experts consider in practice. We also surface a series of tensions that emerge from the analysis. These tensions can guide ethical reflection and raise interesting questions for further investigation. We argue that there is more to ethical guidance in data visualization than a set of rules can provide. Ethics is an ongoing practice demanding the development of good judgment through time, experience, and focused attention. The foundation we offer in this work of a vocabulary of values and set of tensions serves to support reflection, conversation, and the continued development of ethical awareness in data visualization research, education, and practice. 




\acknowledgments{%
This research is sponsored in part by the U.S. National Science Foundation through grant SMA-2418908. We also thank John Basl, Melanie Tory, and Sydney Purdue for their feedback on this work, and our interview participants for their willingness to be open and candid about their experiences.


}

\bibliographystyle{abbrv-doi-hyperref}

\bibliography{vis-ethics}


\end{document}